# Temperature Tunable Optical Transmission control of VO$_2$ nanostructures by IR based 1-D Photonic crystals as hybrid Photonic absorbers

**Dipti Umed Singh*[1], Omkar Bhoite [1] and Remya Narayanan*[2]**

[1]*Department of Physics & Centre for Energy Science, Indian Institute of Science Education and Research, Pune 411008, Maharashtra, India*

[2]*Department of Environmental Science, Savitribai Phule Pune University, Pune, 411007, Maharashtra, India*

***Corresponding authors' emails:** remyapadmaja@gmail.com ; dipti@students.iiserpune.ac.in

# Abstract

Effect of 1-D photonic crystals on optical transmission of $VO_2$ is studied by depositing thin films of $VO_2$ nanoparticles on $SiO_2/TiO_2$ distributed Bragg reflectors (DBR) in the near infrared (IR) spectrum as per earlier theoretical predictions of *J. Phys. D: Appl. Phys. 51 375102 (2018)*. Monoclinic $VO_2$ nanoparticles with tuned crystallinity were synthesized by a facile solution processing method. Moderately crystalline (MC) and highly crystalline (HC) $VO_2$ nanostructures were obtained by varying its synthesis temperature and post growth annealing conditions. Both MC $VO_2$ and HC $VO_2$ films exhibit expected reduction in optical transmission in the IR region due to its structural phase transition from monoclinic (insulator) to rutile (metallic) around critical temperature of 68 °C. By combining $VO_2$ films on a 40% transmitting DBR structure, the average optical transmission further went down to ~ 20%. Number of stacks of DBR plays a key role in such effective reduction of optical transmission in IR. When the number of stacks of DBR is further increased from 4 to 7, optical transmission of metallic $VO_2$ films on DBR nearly vanishes in Near-IR spectrum in such vanadium dioxide/1D photonic crystal based composite photonic structures. Such temperature controlled, enhanced, broad band optical response can be a promising design for $VO_2$ nanoparticle based hybrid photonic absorbers for various smart window applications.

# 1. Introduction

Monoclinic Vanadium dioxide $VO_2$ (M) is a prototype material which can undergo fully reversible metal to insulator transition (MIT) above a critical temperature ($T_c$) of 68 °C. This phase transition occurs due to crystallographic phase change from monoclinic M (insulator or semiconducting) phase to rutile R phase (metallic).[1,2] Owing to this, $VO_2$ exhibits predominant changes in its correlated optical and electrical properties.[3,4] Such MIT transition also leads to abrupt changes from IR transparent semiconducting M state to infra-red (IR) blocking metallic R state,[5] with resistance changing by few orders of magnitude.[6] Such intriguing properties make $VO_2$ as a promising candidate for active optoelectronic and infrared memory devices [7] and this has been extensively used for smart window coating and IR based optical absorbers.[8,9] $VO_2$ based metamaterials are also designed for broad band tunable terahertz absorber.[10-12]

On the other hand, 1-D photonic crystals are one of the most important yet, simple artificial structures to accomplish optical control by localizing light modes and by controlling the flow of light. [13,14], which consists of one dimensional periodic array of multilayers of two materials with different dielectric constants yields very high reflectivity called as Distributed Bragg reflector (DBR). DBR based Photonic crystals are already used as a back reflector for enhanced emission from heterostructure LEDs [15] and also as a directly absorptive medium.[16-18] Such structures also play key roles in wider applications of thermos-photovoltaics and IR photodetectors for energy conservation.[19-22] There were widespread applications of DBR in Bio-Sensing devices[23,24] as well chemical sensors[25,26] and IR imaging Device.[27] 1D Photonic crystal based hybrid structures were studied intensively in order to achieve perfect absorption due to localized photonic modes.

Zhu et. al recently reported usage of an anti-reflection coating in order to enhance the visible transmittance of $VO_2$ film.[28] There are reports of theoretical predication of absorption enhancement using this type of nanocomposite photonic structure on DBR as well.[29,30] Here we have used a 1D photonic crystal based DBR, fabricated with $SiO_2$/$TiO_2$ alternating layers along with $VO_2$ (M) nano-composite thin films. We observed significant reduction of optical transmission in the IR regime.

# 2. Experimental Details

## 2.1 Material Used

Ammonium meta-vanadate ($NH_4VO_3$, 99%, Sigma-Aldrich), ethylene glycol ($C_2H_6O_2$, 99.8%, Sigma-Aldrich), Polyvinylpyrrolidone (PVP, Sigma Aldrich), Ethanol, Silicon dioxide ($SiO_2$) and titanium dioxide ($TiO_2$) sputtering targets of two inch diameter, 0.25 inch thickness were used to fabricate the DBR on double side polished glass slides.

## 2.2 Synthesis of $VO_2$ nanoparticles

1 g of $NH_4VO_3$ was added to 50 ml of ethylene glycol and this reaction mixture was kept at 160 °C in an oil bath for 2 hours with vigorous stirring, resulted the formation of purple color Vanadyl ethylene glycol (VEG) complex. This VEG complex was allowed to cool down to room temperature. After cooling, the complex was centrifuged several times with copious amount of ethylene glycol and ethanol to remove residues from the complex precipitate. Obtained precipitate was further dried for 4 hours at 60 °C and then heated at 190 °C for 30 minutes in a vacuum oven in an ambient atmosphere. This leads to the formation of black power of $VO_2$, which is moderately crystalline (MC). This MC $VO_2$ powder was further annealed at 650 °C for 2 hours with a ramp rate of 5 °C/minute in an Argon air flow. Before annealing, tube furnace was purged with argon gas for 30 minutes to remove the excess oxygen from the chamber to avoid oxidation of $VO_2$ powder. The system was allowed to cool down to room temperature leading to the formation of high crystalline $VO_2$ (HC) powder.

## 2.3 DBR Fabrication

$SiO_2/TiO_2$ multilayers were grown using Moorfield's Minilab ST80A magnetron Sputtering deposition system on thoroughly cleaned microscope glass slides of size 0.5x0.5 cm$^2$. The base pressure of a sputtering chamber was kept at ~$1\times10^{-7}$ mbar. While depositing $SiO_2$ and $TiO_2$, the vacuum chamber was kept at $9.2\times10^{-3}$ mbar. Thickness of $SiO_2$ and $TiO_2$ were calibrated using a Profilometer (Veeco, Dektak 150). Two DBRs with central wavelength of $\lambda_0 = 1350$ nm and 1600 nm were fabricated respectively. For DBR with central wavelength $\lambda_0 = 1350$ nm, four periods of $SiO_2/TiO_2$ was used. Thickness of each layers as $t_{SiO2} \sim 230$ nm and $t_{TiO2} \sim 130$ nm were chosen to be around d = λ/4n, where λ is the central wavelength, n is the refractive index of the material used with refractive indices $n_{SiO2}$ – 1.45 and $n_{TiO2}$ – 2.16 respectively. For DBR with central wavelength $\lambda_0 = 1600$ nm, seven pairs of $SiO_2/TiO_2$ with required thickness of $t_{SiO2} \sim 275$ nm and $t_{TiO2} \sim 150$ nm were grown similarly.

**2.4 Characterization Techniques**

Morphology of $VO_2$ nanostructures were investigated by field emission scanning electron microscopy (FESEM, Zeiss Ultra Plus) and by transmission electron microscope (JEOL JEM-2200FS). High resolution transmission electron microscopy (HRTEM) images and selected area diffraction (SAED) patterns were obtained by JEOL – 2200FS operating at 200 KV. Chemical compositions were analyzed by Oxford's 80 $mm^2$ EDX detector. High temperature X-ray diffraction pattern was recorded by Bruker D8 Advance X-ray diffractometer with Cu Kα radiation (1.54 Å) with temperature ranges were from 300 K to 380 K. The two theta scanning range was from 10 to 70°. Thermochromic properties of $VO_2$ were explored with a homemade setup consisting of Acton Research's SP2555i monochromator to scan wavelength range from 900-2500 nm. The samples were kept inside customized copper sample holders inside ARS CS204-DMX-20 closed cycle cryostat for temperature dependent measurements. The temperature was varied from 300 K to 380 K. Finally, the transmitted and reflected light through the sample was detected with Thorlab's FGA20 InGaAs photodetector.

## 3. Result and discussions

**3.1 structural and morphological analysis of $VO_2$**

VEG complexes were obtained by the reduction of $V^{5+}$ in $NH_4VO_3$ to $V^{4+}$ by pyrolysis in ethylene glycol. [31] This provides a planar sheet like morphology as shown in Figure S1(a).

The average length of each sheet is of a few micrometers. Thermolysis of VEG complexes leads to the formation of MC $VO_2$ in ambient atmosphere, according to the following equation.

$$2VO(OCH_2CH_2O) + 5O_2 \rightarrow 2VO_2 + 4H_2O + 4CO_2 \quad \ldots\ldots\ldots\ldots\ldots\ldots (1)$$

FESEM images showing rod like morphology for as prepared MC $VO_2$ in figure 1(a). TEM images further confirm formation of nanorods with a size distribution ranging from 100 – 200 nm and these are aggregated in clusters.

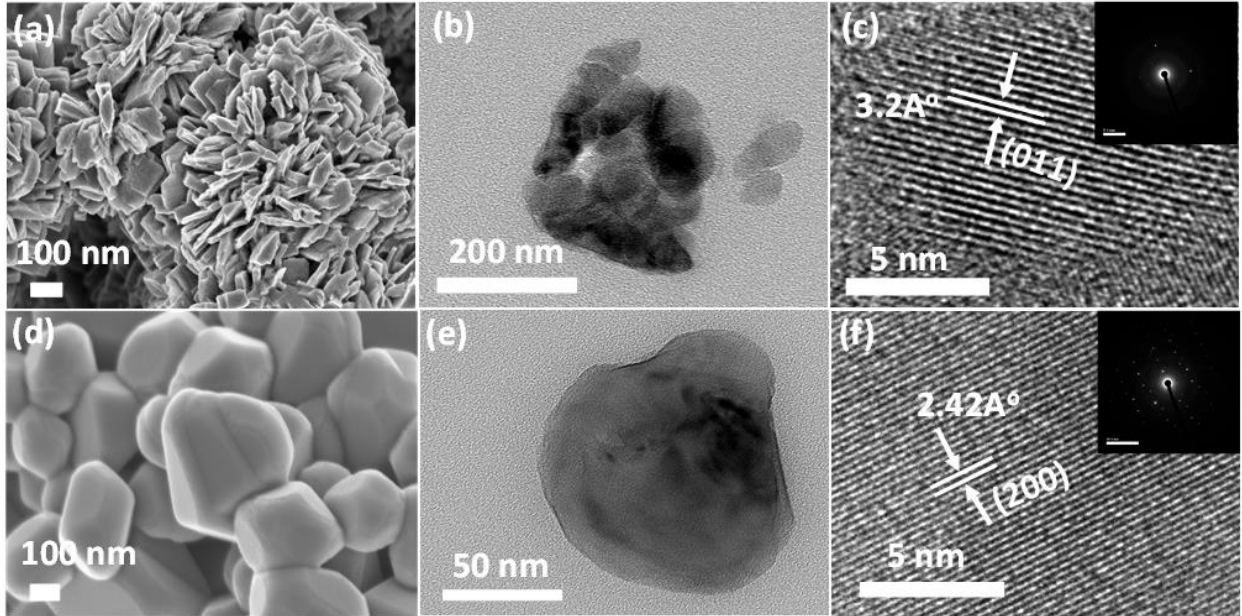

**Figure 1.** (a) and (d) showing the FESEM images of moderate and high crystalline $VO_2$ and (b) and (e) are TEM images of clusters of MC-$VO_2$ nanostructures and regular shaped polyhedron of HC-$VO_2$, (c) and (f) exhibits HRTEM images MC and HC $VO_2$ nanostructure with inset of SEAD pattern

It is assumed that, VEG complexes are acting as seed for $VO_2$ nanorods. The surface energy of $VO_2$ nanorods formed at ambient temperatures is relatively large and they can aggregate well together. Figure 1(c) depicts the high resolution TEM. This provides the information about the $VO_2$ crystalline phase. The calculated inter-planar distance of $VO_2$ nanocrystals is around 3.2 A°, this corresponds to (011) plane of MC-$VO_2$ (M) (JCPDS: 82-0661). Post annealing of as prepared MC $VO_2$ at 650 °C resulted in the formation of HC $VO_2$ nanoparticles with uniform distribution. This morphology is consistent with previous reports in the literature. As formed HC $VO_2$ has regular and clear grain boundaries as shown in figure 1(d). Xiao et. al. [32] reported that, when annealing temperature is increased, the surface energy of the particles with large crystalline size is reduced and this leads to reduced aggregation.[32] The calculated inter-planar distance of HC $VO_2$ (M) polyhedron is 2.42 Å which corresponds to the (200) miller indices.

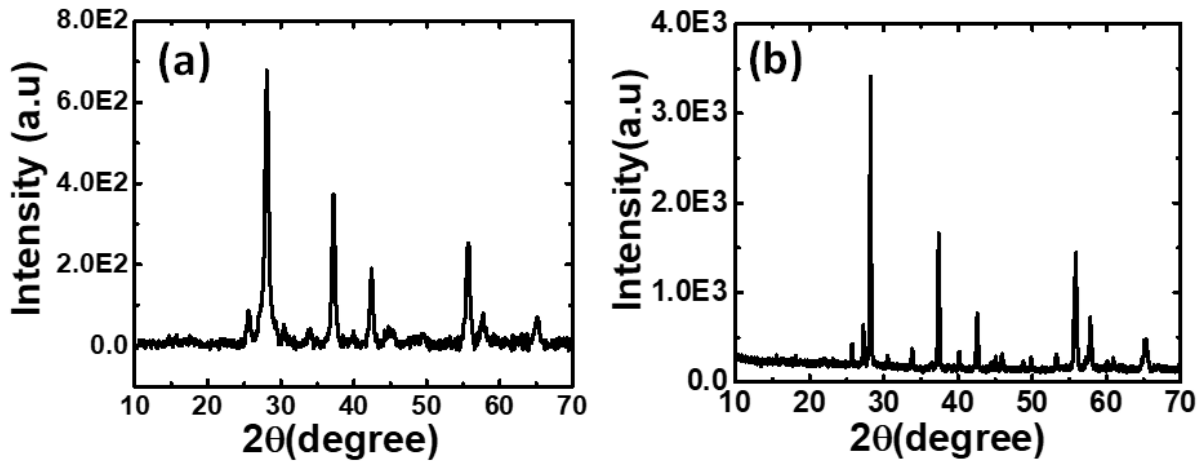

**Figure 2.** XRD of (a) moderately crystalline $VO_2$ formed by annealing at 190 °C in ambient atmosphere and (b) highly crystalline $VO_2$ nanostructures formed by annealing MC $VO_2$ at 650 °C in argon atmosphere.

Figure 2(a) and 2(b) clearly shows that the as-prepared $VO_2$ in both moderate and high crystalline structures were pure for Metal to Insulator (MIT) transition. XRD pattern of as-prepared VEG precursor, with monoclinic phase [JCPDS-49-2497], has highest intensity peak located at $2\Theta = 13.46°$ in Figure S1(b). After thermolysis, VEG peaks started to disappear and started to form MC-$VO_2$ monoclinic phase with preferred orientation along (011) plane at $2\Theta = 27.95$ ºC. As the post annealing temperature is increased further, the XRD peak intensity of $VO_2$ nanoparticles enhanced considerably with decreased full width half maximum (FWHM), thus obtained HC-$VO_2$ as shown in Figure 2(b).

Temperature dependent in situ XRD in Figure 3 confirmed the all-important structural phase transition from monoclinic $VO_2$ (M) to rutile $VO_2$ (R) phase. Previous reports show that, below the transition temperature of 360 K, XRD pattern matches with (JCPDS-82-0661) data confirming the monoclinic phase, whereas above 360 K rutile phase (JCPDS-79-1655) was confirmed. [33] Shifting of the highest peak in XRD from $26º \leq 2\Theta \leq 29º$ in figure 3(a) was a clear indication of such structural phase transition in which $VO_2$ (M) (011) transient into $VO_2$ (R) (110). Also in Figure 3(c), in the range of $64º \leq 2\Theta \leq 66º$, we can observe some splitting of $VO_2$ (M) (310) peak into two peaks of $VO_2$ (R) (130) and $VO_2$(R) (002). The Same structural transition can be seen in high temperature XRD of HC $VO_2$ (M) with highest peak shift varies from $26º \leq 2\Theta \leq 29º$ in Figure 4(a).

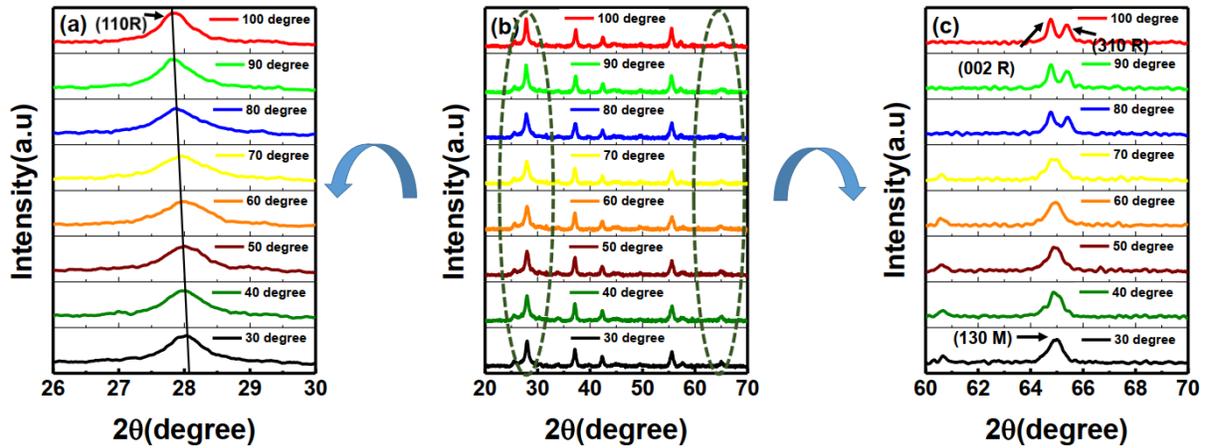

**Figure 3.** High Temperature XRD of MC-VO$_2$. (b) showing the high temperature XRD of moderately crystalline VO$_2$ in both below and above the metal-insulator transition from VO$_2$ (M) to VO$_2$ (R) above 68 °C (a) showing the shift in highest intensity peak in range of 26-30° with temperature (c) showing the gradual splitting of two peaks in range of 62-68° from a single peak at 65°. This confirms the transition of (130) monoclinic plane into two (002) and (310) Rutile planes.

Two sharp, distinct peaks were also visible in range from 64° ≤ 2Θ ≤ 66° in Figure 4(c) above the transition temperature.

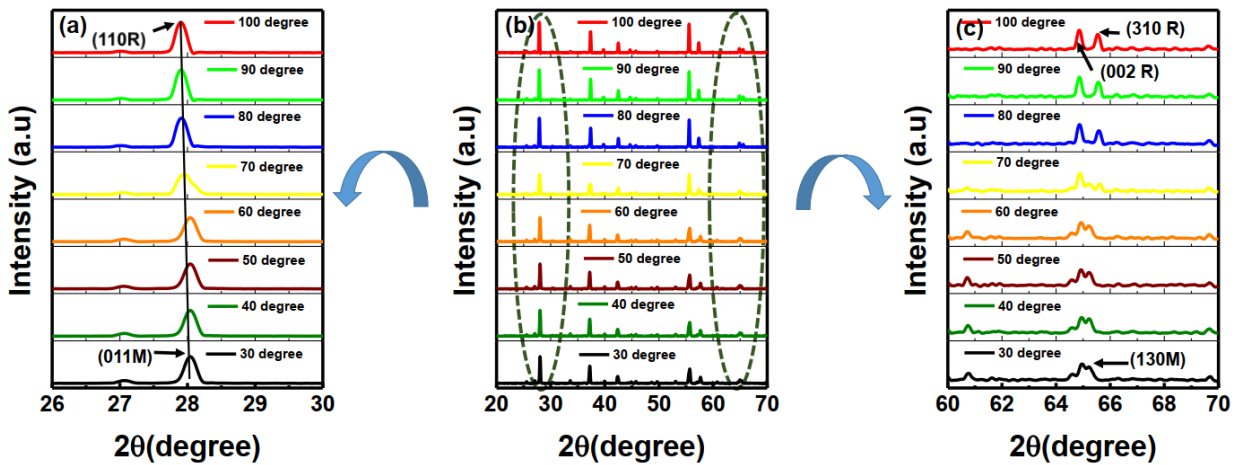

**Figure 4.** High Temperature XRD of HC-VO$_2$. (b) Showing the high temperature XRD of highly crystalline VO$_2$ undergoing the transition from VO$_2$ (M) to VO$_2$(R) above 68° C. On the left, (a) showing the shift in highest intensity peak with temperature. (c) Showing similar gradual splitting of two peaks in range of 62-68° from a single peak at 65°.

Therefore, temperature dependent XRD establishes the structural phase change following the change in dimension of the unit cell from monocline VO$_2$ (M) to rutile VO$_2$ (R). The vanadium atoms with zig zag type atomic orientations in VO$_2$ (M) results in its characteristic insulating behavior. This shifts to the linear chains of vanadium atom to form $VO_2(R)$ during the structural transformation above 68°C.

### 3.2 Optical Characterization of the Composite Photonic Structure

The optical transmission spectra of highly crystalline (HC) and moderately crystalline (MC) $VO_2$ thin films are shown in Figure. 5 (a) and 5 (b). It has been observed that both MC and HC $VO_2$ films were showing similar kind of transmission reduction with increase in temperature. Both the films were fabricated by drop casting over glass slides and equal amount of $VO_2$ were taken in all cases. The change in optical transmission is prominent when the temperature increased above 360 K. For MC $VO_2$ film, with increase in temperature from 300 K to 380 K, optical transmission drops from 67 % to 53 % at 1300 nm and from 73 % to 55 % at 1600 nm. The calculated optical transmission modulation is ($\Delta T$) ~ 14 % at 1300 nm. Similarly, HC $VO_2$ exhibited a similar $\Delta T$ ~10 % at 1600nm when the temperature varies from 300 K to 380 K.

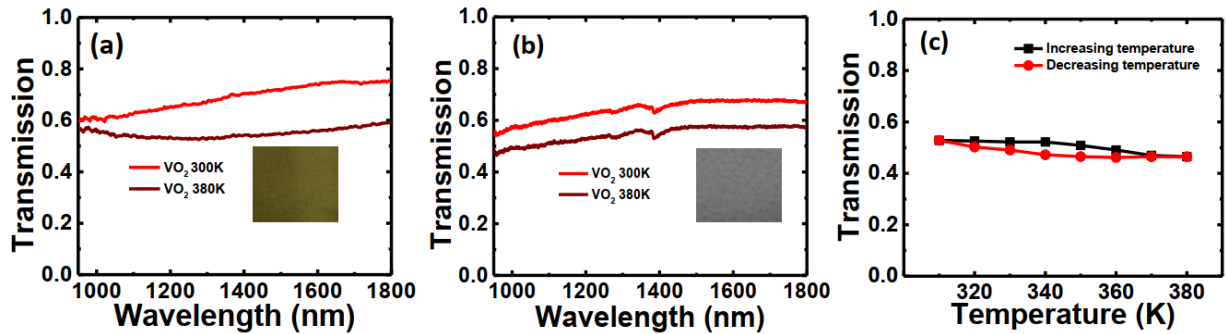

**Figure 5** (a) Showing the transmission spectra of moderately crystalline $VO_2$ film over glass with temperature varying from 300 K to 380 K having transmission change of ~14 % (b) Showing the optical transmission of highly crystalline $VO_2$ nanostructures over glass with temperature variation from 300 K to 380 K having nearly same transmission change of about ~ 10% similar as that of MC $VO_2$ (c) Showing the transmission reversible thermal hysteresis of MC-$VO_2$ film with 0.05 width around critical temperature of 340 K.

The phase structure of $VO_2$ thin film changed effectively from monoclinic to rutile above the critical temperature. Similar transmission reduction is reported by Xygkis. et.al [34] where transmission reductions of ~ 10 % for $VO_2$ film at 1300 nm, above and below the critical temperature is observed. Figure 5(c) showing the transmittance reversible thermal hysteresis of MC-$VO_2$ (M) film over glass at fixed wavelength of 1600 nm. Hysteresis width of MC-$VO_2$ (M) film which is around 0.05 around critical transition temperature of 340 K. Narrow width of hysteresis loop attributed to smaller granular film and good crystallinity of MC-$VO_2$ as reported [35-36]. As both MC and HC are capable of delivering a comparable transmission reduction around the phase transition, hereafter for further studies, we employ MC VO$_2$, which provides the advantage of ease of having a synthesis protocol at ambient atmosphere itself.

### 3.2.1 The Fabry-Perot type temperature dependent optical effect

The hybrid structure with VO$_2$ on 1D Photonic crystal is fabricated as shown in schematic Figure 6(a). Figure 6(b) showing the optical transmission of 40 % transmitting DBR formed with only four periods of SiO$_2$/TiO$_2$ and tuned at central wavelength $\lambda_o \sim$ 1350 nm. On depositing VO$_2$ layer over DBR, it decreases further to 30% with a slight shift of 60 nm i.e, from 1350 nm to 1410 nm towards higher wavelength with ~25 % decrease in optical transmission as compared to film on glass. With increase in temperature, VO$_2$ transmission drops downs further to ~20 % at 380 K within the broad wavelength range of 1250-1550 nm. A few periods of Bragg reflector stacks are therefore capable of delivering the desired amount of reduction in optical transmission. Such VO$_2$/DBR based hybrid optical absorbers can be effectively used for various thermochromic applications.

Consecutively ~100 %, reflecting DBR was also fabricated by increasing the no. of pairs of SiO$_2$/TiO$_2$ stacks as per equation (2). Here, increasing the number of alternate layers in DBR can increase the reflectivity or by increasing the refractive index contrast between the two alternating layers can increase both reflectivity and bandwidth of the reflector. By increasing no. of layers results not only in reduction of transmission modulation but also relativity sharper transmission stop band (Fig.S2) which can be a potential design for other optoelectronic devices with enhanced monochromaticity. Therefore, a DBR with nearly 100 % reflectance is fabricated by increasing the no. of stacks of SiO$_2$/TiO$_2$ at central wavelength $\lambda_o \sim$1600 nm from four to seven and it has only ~10 % optical transmission as demonstrated in Figure 6(c). Therefore, by increasing the no. of layers in the 1D photonic structure one can get nearly 100 % reflectance i.e. effectively a near unity absorption at a desired wavelength.

$$R=[n_o(n_L)^{2N} - n_s(n_H)^{2N} / n_o(n_L)^{2N} + n_s(n_H)^{2N}]^2 \quad \ldots\ldots\ldots\ldots..\ldots\ldots\ldots\ldots\ldots(2)$$

When the temperature of VO$_2$/DBR (~ 5% transmitting) increased to 380 K, its optical transmission further decreased to 0.6 % with a slight spectral red shift to 1668 nm. Thus, it is possible to get almost zero optical transmission or effectively 100 % absorption of VO$_2$ films. This clearly indicates that the number of stacks (n) of DBR plays very important role in such

enhancements. In Figure 6(c), near the stop band, the optical transmission decreases with increase

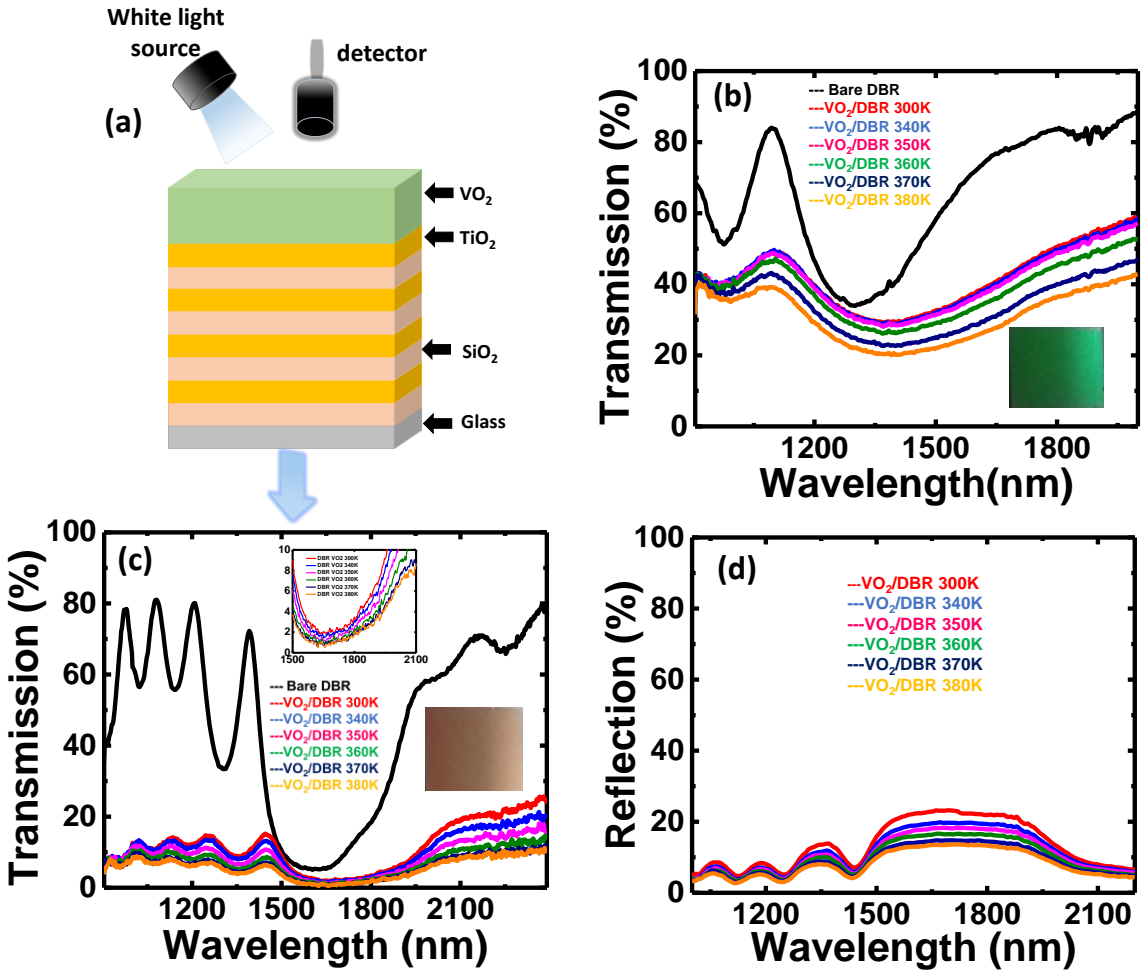

**Figure 6** (a) Showing the schematic diagram of $VO_2$/DBR structure with white light source illumination (b) showing the transmission spectra of MC $VO_2$ film over four stack DBR with a central wavelength tuned at $\lambda_\circ$ ~1300nm and has ~ 40 % transmission. There is clear overall decrease in optical transmission from 34% to 20% with slight red shift with increase in temperature from 300 K to 380 K (c) showing the Transmission spectra of $VO_2$ over seven stack DBR (DBR has ~ 10 % transmission) at $\lambda_\circ$ ~ 1580 nm, with increase in temperature from 300K to 380K transmission decreased to 0.6 % with slight red shift. Inset showing the clear transmission decrease with increase in temperature with constant transmission fractional change $T_{380K}/T_{300K}$ ~ 2% over whole wavelength range and (d) showing the reflection spectra of $VO_2$/DBR with increasing temperature.

in temperature. However, this not as clearly visible in regions below and above the stop band from the given plot. But the fractional change in transmission $T_{300K}/T_{380K}$ in the overall range of wavelengths remains nearly same around 2 %. Inset of Figure 6(c) clearly indicates the change in optical transmission with temperature. Fig 6 (d) is demonstrating the reflection spectra of same

VO$_2$/DBR film of which the transmission spectra shown in Fig 6 (c). VO$_2$/DBR exhibits ~ 30 % reflection at 300 K, on increasing the temperature reflection further decreases to ~ 15 % in IR region. These results clearly indicate that there is an effective absorption enhancement of VO2/DBR film, assuming insignificant scattering losses. The slight red spectral shift in both Figure 6 (b) and 6 (c) may be due to the addition of extra layer of VO$_2$ (M) over the DBR. This affects the overall optical thickness of the DBR and causes its optical interface pattern to shift slightly to higher wavelength. [37] Another possible reason of such red shift could be the non-uniformity of the VO$_2$ surface, due to which the effective angle of refraction of light passing through VO$_2$ layer to DBR deviates towards the lower angles from the normal while entering into DBR as shown in the schematic representation Figure 6(a). since light entering in DBR from VO$_2$ is towards a lower angle rather than normal can cause the red spectral shift which is shown by the Matlab simulation program [38] based on the transfer matrix method to calculate the spectral coefficients of this multilayered structure, as per Eq. (3) represents the reflection and transmission coefficients of a DBR can be obtained by Fresnel expressions,

$$R_s = [(n_o - n_1)/(n_o - n_1)]^2, \quad T_s = [4n_o n_1/(n_o + n_1)^2] \quad \ldots\ldots\ldots\ldots\ldots\ldots(3)$$

Where $R_s$ is the reflection coefficient and $T_s$ is the transmission coefficient. If $M_T$ represents the product of transfer characteristic matrices and $M_i = 1, 2$ represents individual layers in the Bragg reflector. And $M_{vo2}$ represents Matrix corresponds to VO$_2$ layer, then final the transfer characteristic matrix is given as,

$$M_T = M_{VO2} * (M_1 * M_2 * M_1 * M_2 * \ldots * M_n)^{q/2} \quad \ldots\ldots\ldots\ldots\ldots(4)$$

Where q is the number of stacks. [39] Calculating this transfer matrix for non-normal angle of incidence shows prominent red shift when light refract towards a lower angle. Figure 7, depicts the simulation results of such red spectral shift in VO$_2$/DBR structure when light incident at lower angles with VO$_2$ refractive index in metallic phase considered to be ~ 2.3 at 1600nm [40] and following equations 3 and 4, here in the program, we are considering that angle dependence

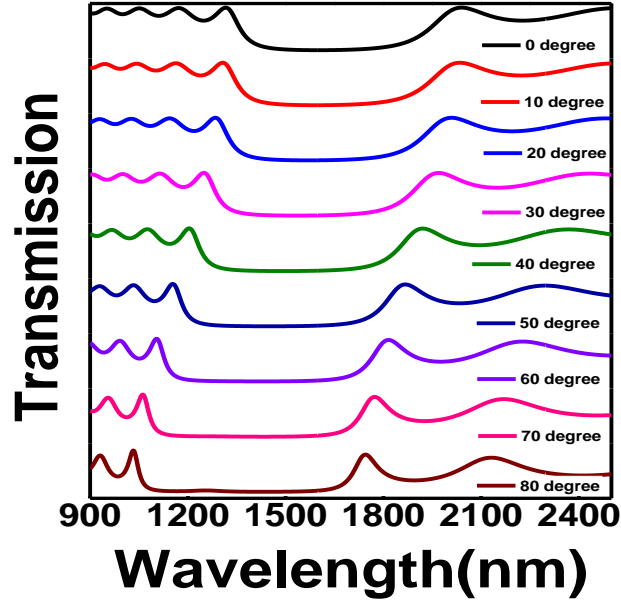

**Figure 7** simulated optical transmission spectra of DBR with varying angle of incidence for s-polarized light transmissivity has negligible scattering loss, thus giving a red shift with a decrease in angle of incidence for s-polarized light in a VO$_2$/DBR structure.

Therefore, it is also likely that some non-uniformity of the VO$_2$ surface may also cause such changes in the effective angle of refraction of light passing from the VO$_2$ thin films into the DBR which causing the red spectral shift of the whole transmission spectra. Optical scattering may also be present due to the non-uniformity of this drop casted film. However, it is difficult to take exact account of optical scattering effects without a full-fledged Kubelka-Munk [41] analyses which is beyond the scope of the present work.

**Conclusion**

In summary, we experimentally demonstrated the effect of 1-D photonic crystals on optical transmission of VO$_2$ nanostructures by depositing VO$_2$ thin films on SiO$_2$/TiO$_2$ based DBR. Therefore, this experimental work confirms the main theoretical prediction of a theoretical earlier paper *J. Phys. D: Appl. Phys. 51 375102 (2018)* on observing near perfect optical absorption in

Near-IR spectrum in such vanadium dioxide/1D photonic crystal based composite nanostructures. In addition, we also report temperature controlled optical transmission based on the metal-insulator transition of $VO_2$.

We synthesize $VO_2$ by thermoylsis of vandyl complex at 190 °C in ambient atmosphere which yields MC $VO_2$ (M) nanostructures. Post annealing treatment at 650 °C under argon atmosphere further results in highly crystalline $VO_2$ (M). High temperature XRD clearly confirms the crystal phase change from monoclinic to rutile above 68 °C for both MC and HC $VO_2$. Both these films exhibit almost similar transmission reductions (~10 %) due to structural phase change above their metal-insulator transition temperature. Therefore, MC $VO_2$ has been employed for further studies due to its ease of synthesis at ambient atmospheric conditions without any post-growth annealing steps. By depositing $VO_2$ over DBR with four periods of $SiO_2/TiO_2$ we get ~20 % transmission change. On increasing the number or periods of $SiO_2$, $TiO_2$ to seven, results in the formation of a DBR having a stop band at 1600 nm with ~ 10 % transmission. This information may prove to be useful while designing such smart windows and to figure out how many layers will be optimum for desired optical effects for a particular application. By depositing $VO_2$ over a good quality DBR leads to ~ 100 % reduction in optical transmission (0.6 % at 380K) and reflection over a certain wavelength region in IR. Hence such hybrid structures are potential candidates for designing $VO_2$ based temperature controlled optical absorbers for various thermochromic applications such as smart windows and IR sensors. Moreover, our results show that it will be technologically simpler to implement because even moderately crystalline $VO_2$ can show substantial reduction in optical transmission. Further work is going on to tailor these hybrid $VO_2$ nanoparticles and ID photonic crystal structure to lower metal-insulator transition temperature of 68 °C towards room temperature with suitable doping for ambient operations.

**Acknowledgements**

RN and Dipti acknowledges funding from the Department of Science Technology, India through INSPIRE faculty scheme (DST/INSPIRE/04/2017/002761). We also acknowledge the support from Department of Science and Technology, India (Research Grant SR/NM/TP13/2016). We also thank Dr. Shouvik Datta, IISER Pune for valuable discussions.